# Structural, Morphological, Optical and Magnetic Property of Mn doped Ferromagnetic ZnO thin film

R. Karmakar, S. K. Neogi, Aritra Banerjee and S. Bandyopadhyay*

*Department of Physics, University of Calcutta, 92 APC Road, Kolkata-700 009, India*

## ABSTRACT

The structural, optical and magnetic properties of the $Zn_{1-x}Mn_xO$ ($0 \leq x \leq 0.05$) thin films synthesized by sol-gel technique have been analyzed in the light of modification of the electronic structure and disorder developed in the samples due to Mn doping. The films are of single phase in nature and no formation of any secondary phase has been detected from structural analysis. Absence of magnetic impurity phase in these films confirmed from morphological study also. Increasing tendency of lattice parameters and unit cell volume has been observed with increasing Mn doping concentration. The incorporation of $Mn^{2+}$ ions introduces disorder in the system. That also leads to slight degradation in crystalline quality of the films with increasing doping. The grain size reduces with increase in Mn doping proportion. The band gaps shows red shift with doping and the width of localized states shows an increasing tendency with doping concentration. It is due to the formation of impurity band and trapping of Mn atoms, which leads to the generation of the defect states within the forbidden band. Photoluminescence (PL) spectra shows gradual decrease of intensity of exitonic and defect related peaks with increasing Mn doping. Defect mediated intrinsic ferromagnetism has been observed even at room temperature for 5at% Mn doped ZnO film. The strong presence of antiferromagnetic (AFM) interaction reduces the observed ferromagnetic moments.

**Corresponding author*: E-mail: sbaphy@caluniv.ac.in

## 1. Introduction

Wurtzite structured ZnO, a wide band gap ($E_g$= 3.3 eV) semiconductor is a potential candidate for optoelectronics devices [1]. Large exciton binding energy (60 meV), strong room temperature luminescence, high electron mobility, good transparency etc are some of the advantages of ZnO [2, 3]. Beside this, Mn doped ZnO ($Zn_{1-x}Mn_xO$) has also been found as one of the most promising material for spintronic devices with ferromagnetism (FM) retained at room temperature [4, 5]. There are continuous efforts to detect the origin of the observed FM in Mn doped ZnO system [6, 7, 8].

Though report shows that origin of magnetism of this types of samples particularly for Mn doped ZnO samples are contradictory [7, 9]. The main objective in synthesizing this type of dilute magnetic semiconductor (DMS) bulk samples or thin films is to achieve intrinsic ferromagnetism within the system. Any type of impurity phase driven FM actually destroys the objective of DMS samples. The reason of observed intrinsic ferromagnetism in DMS samples [particularly for Mn doped ZnO] is the primary concern of debate. However recently at least a general consensus has attained that defect mediated intrinsic ferromagnetic DMS samples are truly capable of serving as a vital component in spintronic devices [5].

Beside the generation of novel magnetic properties, the introduction of $Mn^{2+}$ ions can also tune the optical properties viz., band gap of ZnO. Generally band-gap is estimated from UV-visible spectrophotometry. But like origin of magnetism, another debatable issue is the nature of variation of $E_g$ with Mn concentration. There are reports of both blue shift [4, 10] as well as red shift [4, 11, 12, and 13] of $E_g$, with increasing Mn in $Zn_{1-x}Mn_xO$ system. In this work an attempt has been made to resolve this controversy and interpreted the observed redshift in Mn doped ZnO film. It has been made on the basis of redistribution of states through defect or disorder introduced in the system through Mn doping.

Another optical measurement viz. photoluminescence (PL) spectroscopy is very crucial for understanding the types of defects present inside the system. Upon excitation with UV light, electrons are pumped from the valence band (VB) to the conduction band (CB). Those electrons rapidly decay to the bottom of the CB by some non-radiative processes. Then it returns to its ground state following different paths, either (a) creation of an exciton and its subsequent annihilation with emitting a photon having energy nearly equal to the band gap energy of ZnO gap or (b) nonradiative transition to some of the intermediate level created by some defects and then subsequent radiative decay to the VB, resulting emission in the visible range [14].

$Zn_{1-x}Mn_xO$ ($0 \leq x \leq 0.05$) thin films were analyzed. Atomic force microscopy (AFM) and X-ray diffraction (XRD) were employed to study the effect of Mn doping on the crystalline quality and morphology of the synthesized films. Band gap has been evaluated from optical absorption spectra and observed red shift has been interpreted. The nature of change of pattern of exitonic as well defect mediated emission peaks with variation in Mn doping has been analyzed by PL spectroscopy. Magnetic property of ZnO film doped with 5at% (maximum) of Mn has been chosen for analysis. Field (at room temperature) and temperature dependent magnetization were analyzed. The evaluated structural, morphological, optical and magnetic parameters have been interpreted in the light of modification of the electronic structure and disorder introduced in the films through Mn doping.

## 2. Experimental details

To synthesize $Zn_{1-x}Mn_xO$ (x = 0.00, 0.01, 0.03, 0.05) thin films by sol-gel technique stoichiometric amount of zinc acetate dihydrate [$Zn(CH_3COO)_2.2H_2O$] and manganese acetate dihydrate [$Mn(CH_3COO)_2.2H_2O$] was added to a solution containing 2-propanol and diethanolamine (DEA). DEA was used as the sol stabilizer. The resultant solution was mixed thoroughly with a magnetic stirrer at room temperature for 3 hours and kept for 48 hours. The precursor solutions thus obtained were used for spin coating on quartz substrates to prepare the films. The spinning rate and period were optimized to 2700 rpm and 35 sec, respectively. After coating, the films were first dried at $300^0$C for 15 minutes in a furnace to evaporate the solvent. It was followed by heat treatment at 500 $^0$C for 30 minutes. The process of coating and subsequent annealing were repeated for ten times to obtain the desired same (within a limit of very small film synthesis in accuracy) film thickness of all the films. To estimate thickness accurately Rutherford Back Scattering (RBS) measurement has been performed for 5at% Mn doped ZnO film.

XRD patterns of the synthesized thin films were recorded with $CuK_\alpha$ radiation using an automatic powder diffractometer (Make-Philips, Model: PW1830), equipped with θ-2θ geometry. All the XRD measurements were carried out in the range of $20^0 \leq 2\theta \leq 80^0$ at a step size of $0.005^0$. The grain size of the synthesized samples was calculated from the XRD pattern using Scherrer's formula [15, 16]. The contribution of instrumental broadening, while estimating the grain size, has been taken into account. The surface morphology of the samples were investigated in AFM and magnetic force microscopy (MFM) modes of multi mode scanning probe microscope (SPM) with Nanoscope III-a Controller (Make-Digital/Veeco Instruments Inc.). The band gap of the thin films was calculated by analyzing the optical absorption spectra obtained by using a UV-Visible spectrophotometer (Make-Perkin Elmer; Model: Lambda 35) in the wavelength range 190-1100 nm. PL spectra were acquired in the wavelength range of 330–600 nm at room temperature; by using a He-Cd laser of 330 nm as an excitation source. The temperature-dependent DC magnetization data was recorded only for the $Zn_{0.55}Mn_{0.05}O$ thin film sample using SQUID magnetometer (MPMS, Quantum Design). The data was recorded in zero field cooled (ZFC) and field cooled (FC) mode in the presence of 200Oe magnetic field in temperature range 300K to 5K. We have also measured the magnetization (M) as a function of magnetic field (H) at room temperature (300K).

## Results & Discussion

The profile of 5at% Mn doped ZnO film emerging from RBS measurement has been shown in Fig. 1. It indicates the profile of the film surface is quite well. The film thickness has been estimated after analyzing the RBS spectra. It has been found to be of (560 ± 10) nm.

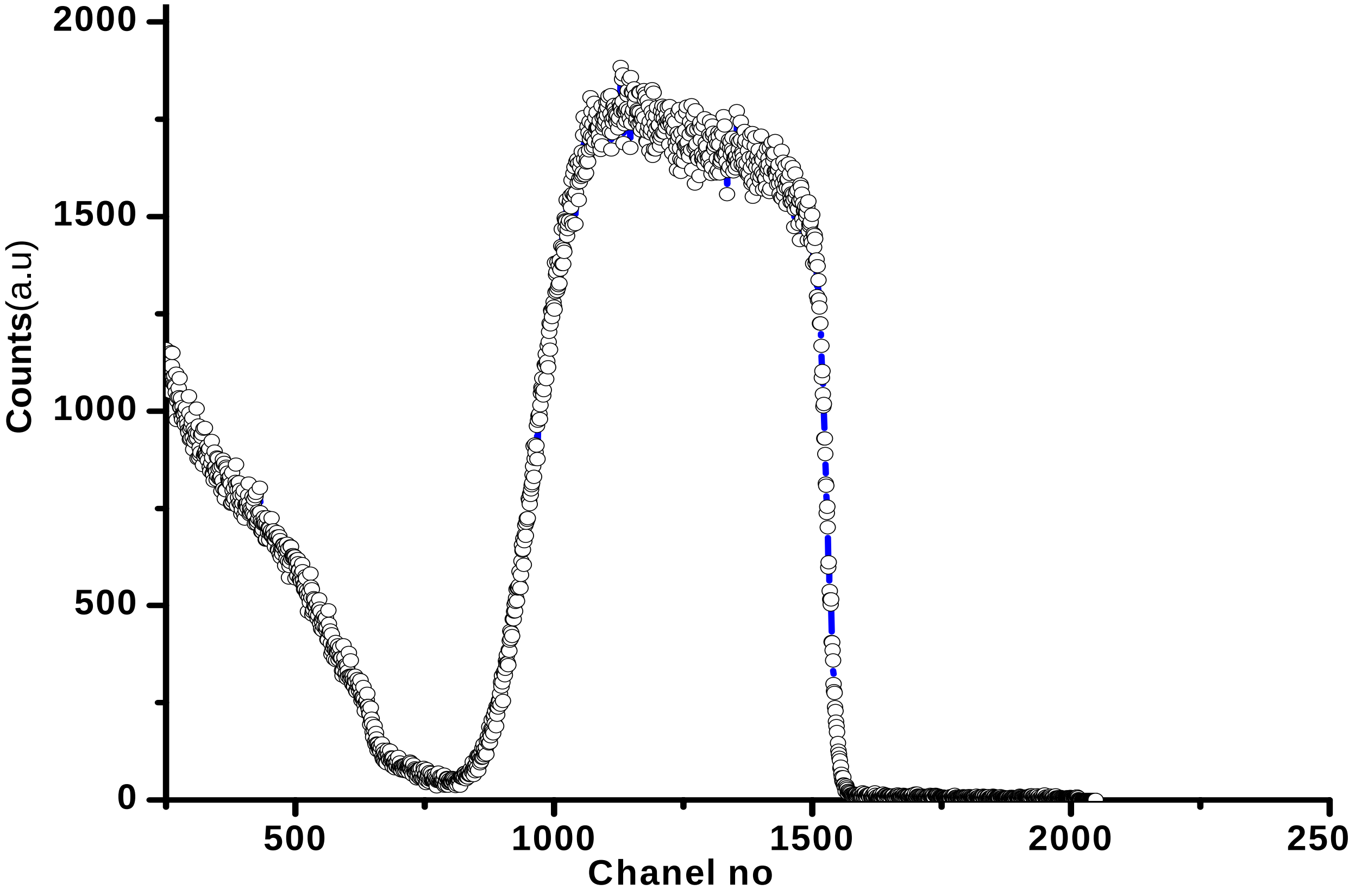

**Fig-1:** *Spectra of RBS measurement for 5at% Mn doped ZnO film. [Thickness estimated from analyzing the spectra is of (560 ± 10) nm].*

XRD pattern of the $Zn_{1-x}Mn_xO$ (x=0.00, 0.01, 0.03, 0.05) thin films deposited on quartz substrate are shown in Fig. 2 (a-d). All of the peaks indexed to a hexagonal wurtzite structure of ZnO, with a dominant (101) peak around $2\theta = 37.5^0$. No diffraction peaks of any secondary or impurity phase were found in any of the thin films. The broad peak around $25^0$ is arising from amorphous quartz substrate as shown in Fig. 2. It has been progressively become prominent from Fig. 2(a) to 2(d). It is due to the fact that with increasing Mn doping gradually the intensities of all the peaks of ZnO are gradually decreasing as shown from Fig. 2(a) to 2(d). So the broad peak arising from amorphous quartz substrate is progressively increasing. Close inspection of Fig. 2(a-d) reveals that, all the XRD peaks including (002) and (101) of the synthesized $Zn_{1-x}Mn_xO$ films were found to shift towards lower angle as Mn concentration increases. The inset of Fig. 2(a-d) shows gradual shifting of (101) peak to lower angle with increasing Mn concentration.

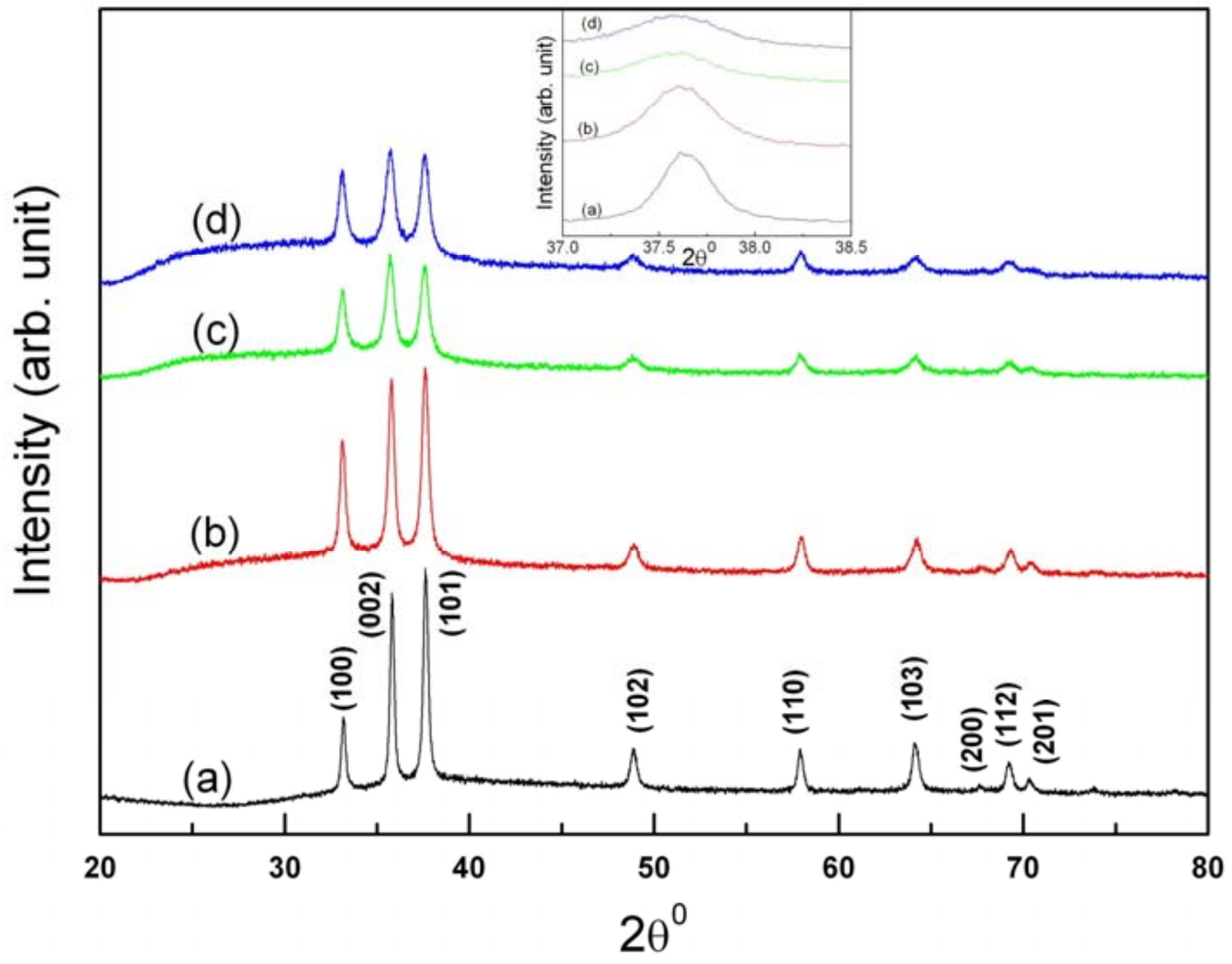

**Fig-2:** *XRD patterns of $Zn_{1-x}Mn_xO$ thin films on quartz substrate (a) x=0.00 (b) x=0.01 (c) x=0.03 (d) x=0.05. Inset shows enlarged view, indicating the lower angle shift of (101) diffraction peaks for above mentioned Mn concentration.*

The shifting of the peak position to lower angle with increase in Mn doping concentration indicates the expansion of the lattice parameters [15]. Increase of c axis lattice parameter with increasing Mn concentration in $Zn_{1-x}Mn_xO$ thin films have already been reported [1, 13, 17]. Also, very recently Deshmukh et al observed a systematic increase of both lattice parameter values viz. *a* and *c* with increase in Mn concentration [18]. The ionic radii of $Mn^{2+}$ ions (0.66 Å) are larger than that of $Zn^{2+}$ ions (0.60 Å) [1, 13]. Thus, such increase of lattice parameter and hence the corresponding expansion of the unit cell volume of the $Zn_{1-x}Mn_xO$ samples with increasing Mn doping is expected. This further confirms that, incorporation of $Mn^{2+}$ ions in the $Zn_{1-x}Mn_xO$ system increases with increasing Mn proportion. Similar result has also been reported in our recent work [17]. Moreover, crystalline quality of Mn doped ZnO films decreases with increasing Mn doping. Actually, the XRD figure (Fig. 2(a-d)) shows that when Mn concentration is increased, intensity of the XRD peak decreases monotonously. Also width of the obtained XRD peaks shows a systematic broadening with increasing doping concentration. These two concurrent observations signify the degradation of the crystalline quality. It is probably associated to the lattice disorder and strain induced in the ZnO lattice due to the substitution of $Zn^{2+}$ ions by comparatively higher ionic radius of $Mn^{2+}$ ions [13].

Fig. 2(e) shows variation of intensity ratios of (100) and (101) peaks, and (002) and (101) peaks with variation in Mn doping proportion. The ratio of intensity of (100) and (101) peaks progressively increases from undoped to 5at% of Mn doping. After 3at% of Mn doping it shows a rather small increase (near saturation). The ratio of intensity of (002) and (101) peaks also shows almost similar kind of behavior. Only difference is that after 3at% of Mn doping it shows small decrease (near saturation). However in the latter case the variation is quite small while the increasing rate with Mn doping is high for former one. (100) peak is progressively emerging with increasing Mn doping. A tendency of increasing polycrystalline nature of ZnO films has been emerged with increasing Mn doping. Finally with 5at% of Mn doping the intensities of (100), (002) and (101) peaks are almost same and it is a quite interesting point to note.

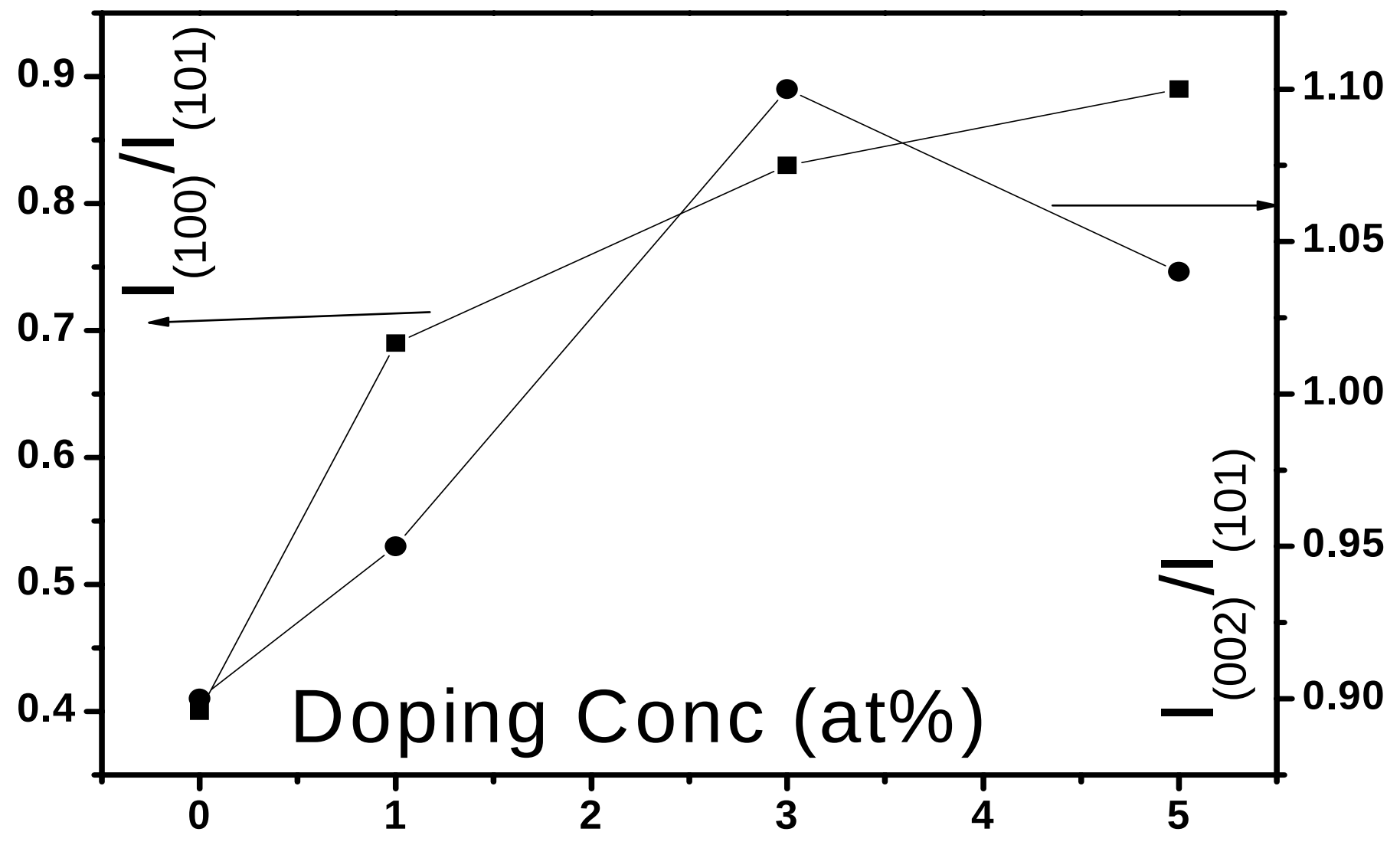

**Fig-2(e):** *Variation of intensity ratios of (100) and (101) peaks, and (002) and (101) peaks with variation in Mn doping proportion.*

The average crystallite size (D) has been estimated from the XRD data using Scherrer's formula [15]. The values of the crystallite size (D), presented in Table-1, indicates a steady decrease with increasing Mn concentration. There exist reports of similar trend in some Mn doped ZnO systems [13, 19]. It is believed that, interstitial zinc plays an important role in the grain growth of ZnO [20, 21]. Actually, the moving boundaries attached the zinc interstitials. Gradual increase of Mn doping progressively reduces the concentration of zinc in the system. Thus the diffusivity is decreased in ZnO, which results in a suppressed grain growth of Mn doped ZnO samples [20]. At the same time, the substituted Mn ions provide a retarding force on the grain boundaries. If the retarding force generated is more than the driving force for grain growth due to Zn, the movement of the grain boundary is impeded [17, 21]. This in turn gradually decreases crystallize size with increasing Mn concentration.

AFM measurement performed to study the surface morphology of the undoped and Mn-doped ZnO thin films. Fig. 3(a)-(d) shows the 2D AFM images. Micrographs reveal that films are closely packed and granular in nature; signature of agglomeration of grains is almost absent.

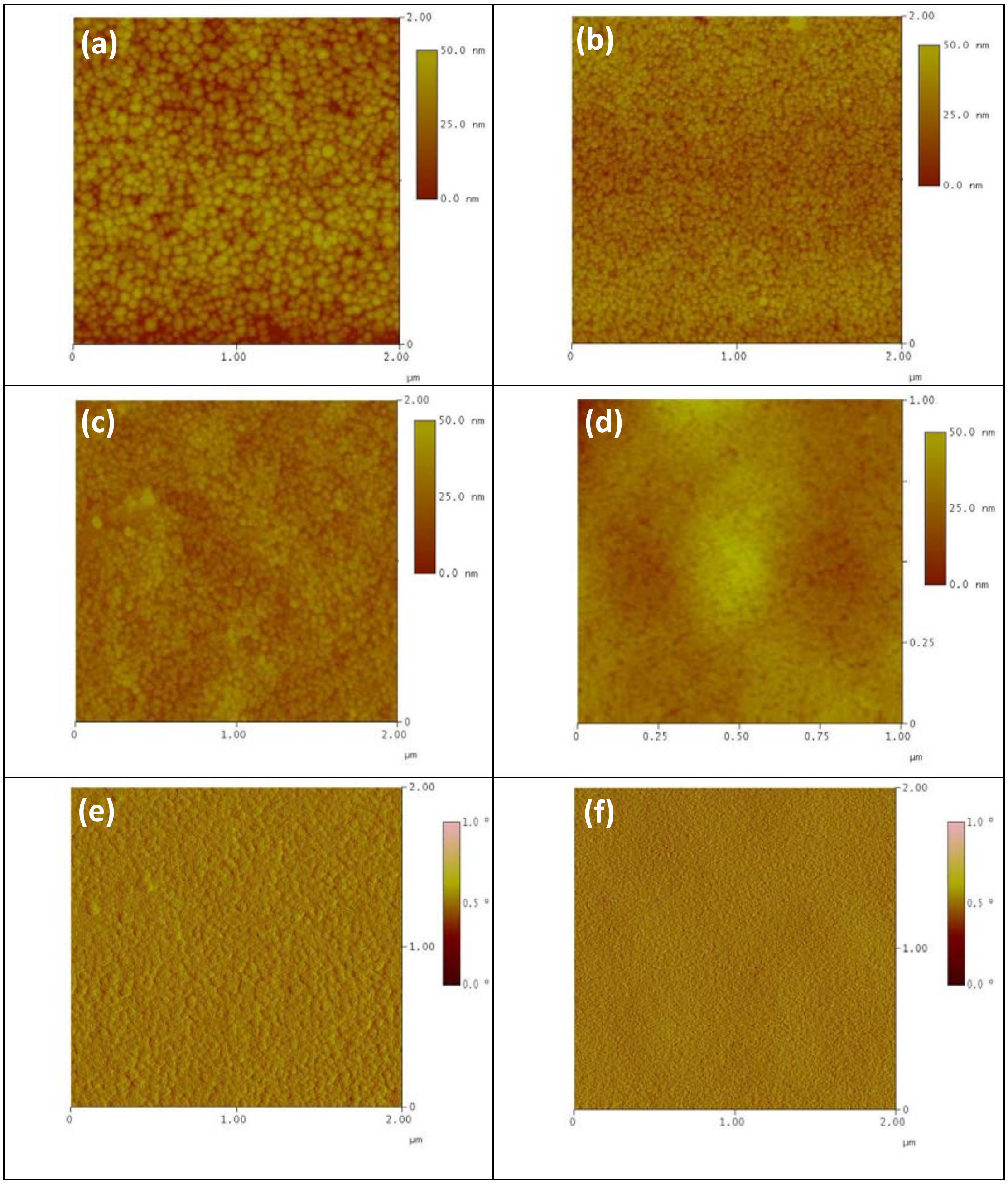

**Fig-3:** *AFM images of undoped and Mn-doped ZnO thin films: (a) undoped (b) 1 at% Mn doped, (c) 3 at% Mn doped, (d) 5 at% Mn doped. MFM images of (e) $Zn_{0.97}Mn_{0.03}O$ and (f) $Zn_{0.95}Mn_{0.05}O$ thin film sample.*

In accordance with the XRD result, close monitoring of the AFM images of films with different Mn concentration further reveal that, crystalline quality of films degrades slightly with increasing Mn doping concentration. AFM images have also been utilized to estimate the grain size of the samples. The estimated values, obtained using *ImageJ*® software, are in close agreement with those obtained from the XRD data (Table 1). It is noteworthy to mention that, similar to XRD results, the grain size estimated from AFM data found to decrease with increase in Mn doping concentration. Fig. 3(e)-(f) depicts the MFM images of $Zn_{1-x}Mn_xO$ (x = 0.03, 0.05) thin films. Uniform brightness contrast exhibits absence of magnetic impurity particles/clusters. Otherwise MFM micrographs with clearer non-uniform brightness contrast would indicate the presence of magnetic impurity in the system [22, 23]. This observation strongly supports the view point of single phase nature of films arisen from XRD spectra.

Optical study of all $Zn_{1-x}Mn_xO$ (x=0.00, 0.01, 0.03, 0.05) thin films has been performed by UV-Visible spectrophotometry PL spectroscopy. UV-Visible spectrophotometry leads to observe the effect of Mn doping on the band gap of the synthesized films. The relation between absorption coefficient (α) and photon energy (hν) is expressed as follows [24, 25]

$$\alpha h\nu = A\,(h\nu - E_g)^{m/2} \qquad (1)$$

Where A is an energy dependent constant, $E_g$ is the optical band gap of the material. m is a constant that depends on the semiconducting materials and m = 1 for a direct transition, whereas for indirect band gap semiconductor m = 4. The linear region of the $(\alpha h\nu)^2$ against hν plot is extrapolated to intersect the energy axis, since $E_g = h\nu$ (direct band gap) when $(\alpha h\nu)^2 = 0$ (Fig. 4). The estimated values of $E_g$ have been presented in Table 1. The obtained value of $E_g$ for undoped ZnO thin film is 3.24 eV. Similar value for thin film of ZnO sample has also been reported earlier [1, 3, 13]. Further, $E_g$ was found to decrease with corresponding increase in Mn concentration (shown in inset of Fig. 4) for these Mn doped ZnO thin films under study.

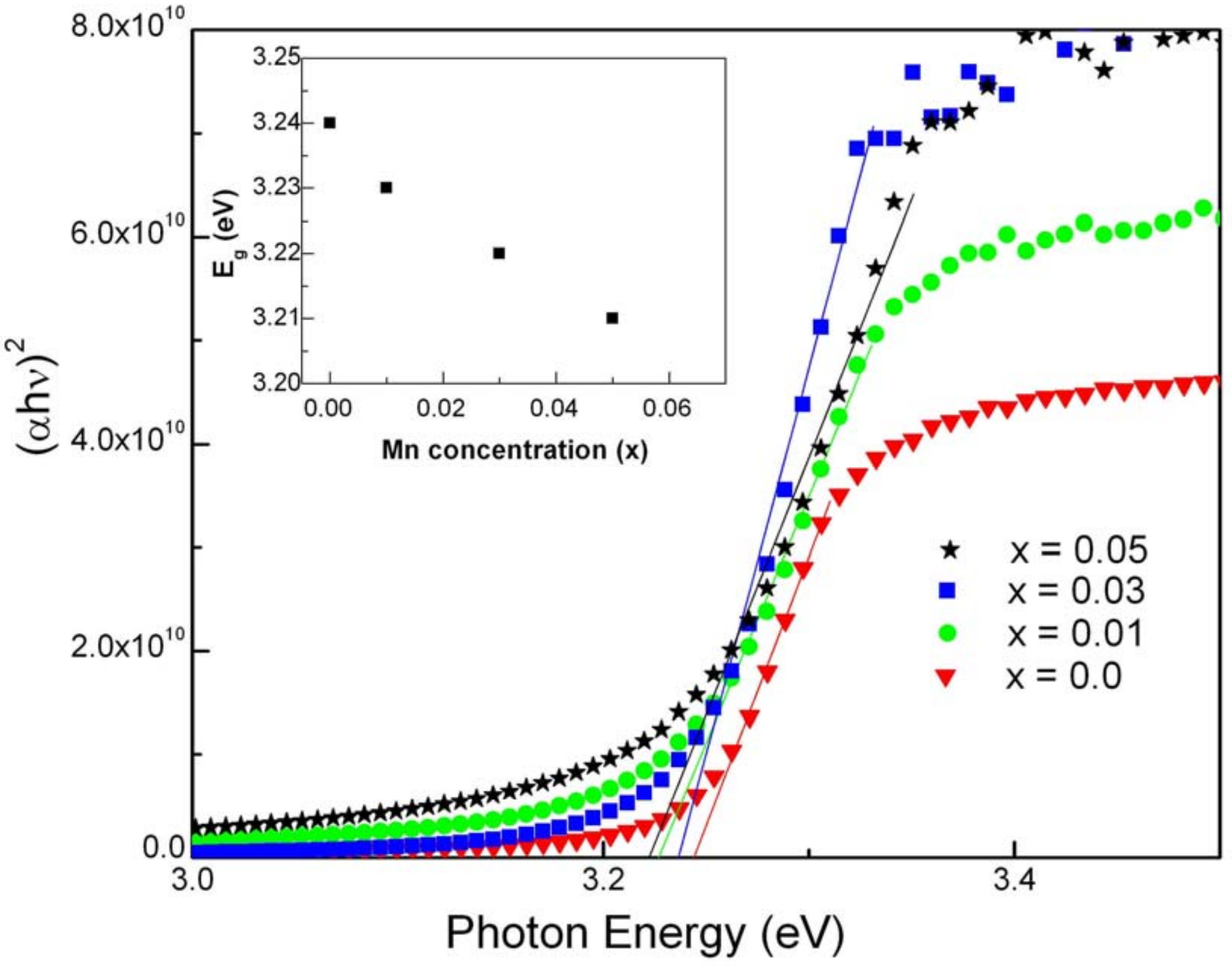

**Fig-4:** *$(\alpha h\nu)^2$ as a function of photon energy for undoped and 1, 3 and 5 at% Mn doped ZnO thin films. The inset shows the variation of band gap ($E_g$) with Mn doping concentration.*

It was observed that, with increasing Mn doping concentration band gap as well as grain size decreases. To explain, a simple model proposed earlier for Al doped ZnO film [26] and bulk Mn doped ZnO [15] has been introduced. According to this model, periodic variations in potential within the grain do occur due to trapping of impurities arising out of doping. Such periodic variation of potential, further leads to the band bending or band tailing effect [13, 24, 26]. Impurity band formation is an obvious consequence of increased doping concentration [25] and the trapping of the Mn atoms at the grain boundary leads to the introduction of the Mn defect states within the forbidden band. With increasing Mn doping, density of this Mn induced defect states increases, leading to the observed decrease of band gap or red shift (Fig. 4(inset)). Similar observation of introduction of Mn defect state within the band gap has also been reported [4, 11, 12]. Actually, trapping of Mn impurities within the grain and the introduction of Mn defect states within the forbidden band gap region is intimately related to the disorder introduced in the system by Mn doping. To confirm this point band gap ($E_g$) has been plotted against inverse of average grain size (1/D) (Fig. 5). 1/D depicts the surface to volume ratio of the constituent particle of the sample. With decreasing grain size more and more disorder is introduced in the system. Thus 1/D can be a measure of disorder introduced in the system. It is quite evident that more disorder should be introduced in the system with increasing Mn doping as ionic radius of the Mn is greater than Zn. Fig. 5 indicates that, with increasing disorder in the system the band gap decreases.

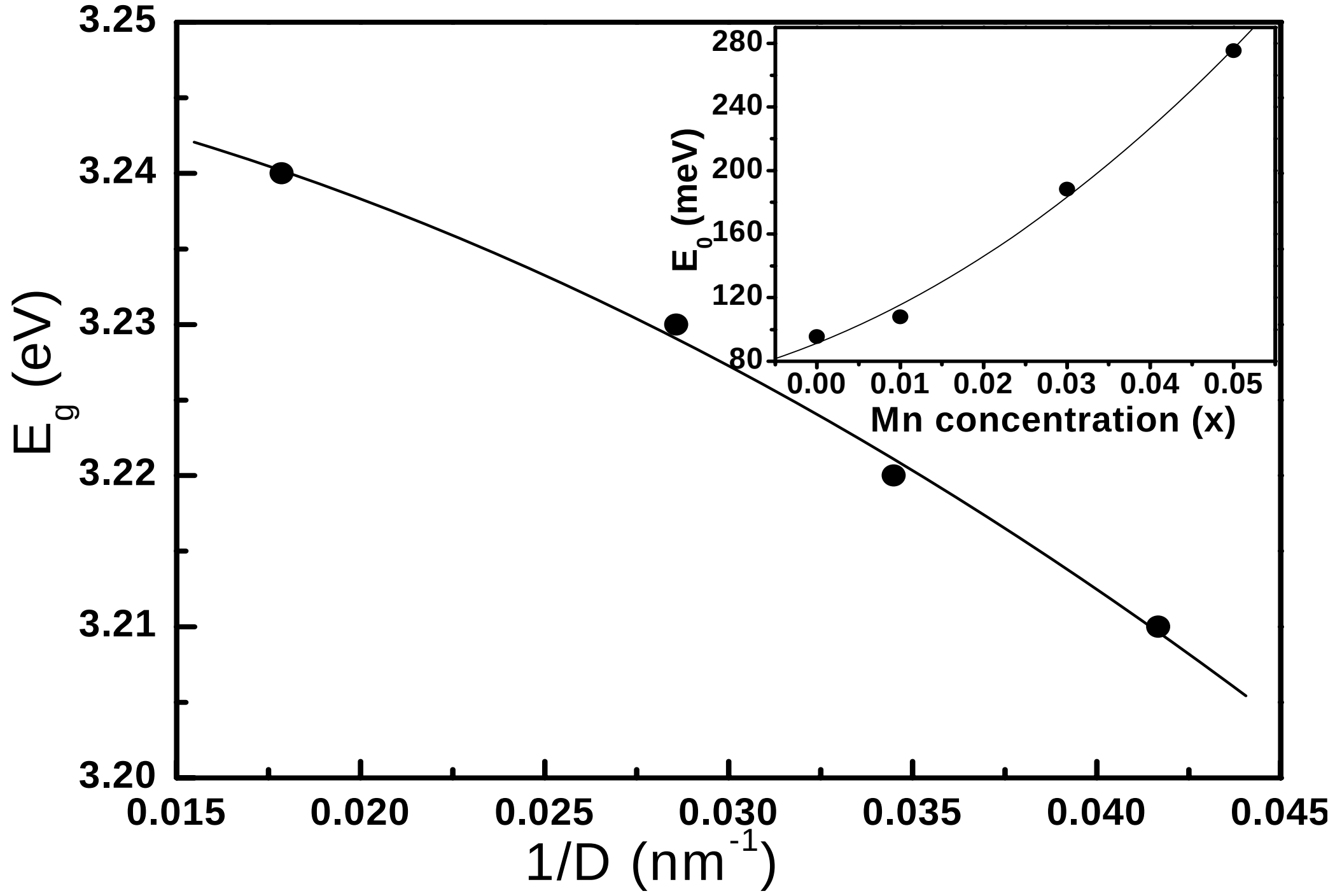

**Fig-5:** *Variation of band gap with inverse of grain size (1/D) for $Zn_{1-x}Mn_xO$ samples. Inset plots the x dependence of band tail parameter ($E_0$) of $Zn_{1-x}Mn_xO$ films.*

The effect of disorder due to doping is more pronounced in lower energy region of band gap, where interaction between valence band holes and/or conduction band electrons and dopant Mn ions leads to bending of bands. This band bending, known as Urbach tail, affects optical band gap structure and hence optical transition [25]. It is noted that both conduction and valence bands can have tail states inside the forbidden region depending on the nature of disorder itself. According to the theoretical calculation [25], the absorption coefficient just below the band edge ($E < E_g$) should vary exponentially with absorbed photon energy (E), i.e.

$$\alpha(E) = \alpha_0 \exp(E/E_0) \tag{2}$$

Where $\alpha_0$ is a constant and $E_0$ is the Urbach tail energy. $E_0$ corresponds to width of localized states arising due to band tailing [25, 27]. Any defect or disorder in the lattice gives rise to a localized potential within the band gap and is directly related to the band tailing parameter $E_0$. $E_0$ have been estimated from reciprocal of the slope of the linear part of ln(α) versus E curve ($E < E_g$) and listed in Table 1.

**Table 1:** *Variation Grain Size, Band gap ($E_g$) and Band Tail Parameter ($E_0$) for $Zn_{1-x}Mn_xO$ thin films.*

| x | Grain size (nm) estimated using | | $E_g$ (eV) | $E_0$ (meV) |
|---|---|---|---|---|
| | Scherrer's formula (from XRD) | *ImageJ*® software (from AFM micrograph) | | |
| 0.00 | 56 | 60 | 3.24 | 95.5 |
| 0.01 | 35 | 38 | 3.23 | 107.8 |
| 0.03 | 29 | 32 | 3.22 | 188.3 |
| 0.05 | 26 | 19 | 3.21 | 275.4 |

The variation of $E_0$ with Mn doping has been plotted in inset of Fig. 5. It shows a monotonic increase of $E_0$ with Mn incorporation. Similar result for ZnO with different kind of dopant has also been observed earlier [3]. The increase in $E_0$ leads to a redistribution of states, and in turn a decrease in the optical gap occurs due to broadening of the Urbach tail. This follows naturally the red shift of the band gap. Though the reduction in band gap values with increasing Mn doping is systematic however it is not so appreciable. But the increase in $E_0$ values with increasing Mn doping is systematic as well as appreciable. Actually the estimation of $E_g$ is rather a fitting dependant where as dependence of fitting is significantly less in the calculation of $E_0$. Interpretation seems to be accurate with more reliability on $E_0$ values. So the effect of reduction of grain size with increasing Mn doping has been responsible for changes in optical parameters.

Fig. 6(a)-(b) shows room temperature photoluminescence (PL) spectra of undoped and doped $Zn_{1-x}Mn_xO$ (x = 0.00, 0.01, 0.03, 0.05) thin films. The little variation PL intensities of all the films indicate a slight variation of film thicknesses. The variation of film thickness is quite random and it has no direct correlation with Mn doping. In the PL spectra a UV emission peak centered at around (~ 380 nm) and a broad peak in the visible emission region with higher wavelength (500-550nm) have been found. In the room temperature PL spectra of undoped ZnO thin film a dominant peak around (~378 nm or 3.27eV) has been observed. The peak in the UV region corresponds to near band edge emission (NBE), because this peak is located close to the band gap energy (~3.3 eV), of ZnO at room temperature [1]. The UV emission is attributed to the radiative recombination of a hole in the valence band with an electron in the conduction band. It is quiet remarkable that UV PL emission peak for all the doped and undoped films was broad and asymmetric in shape and bears a shoulder on the lower energy side. The origin of such a broad transition in the NBE region has not been resolved completely. Several possibilities like electron transition to shallow acceptor states [28], donor bound exciton transition [29] or combination of free exciton transition along with its phonon replicas [30] can be responsible for such broad transition. It has been observed that UV emission is stronger in undoped ZnO sample.

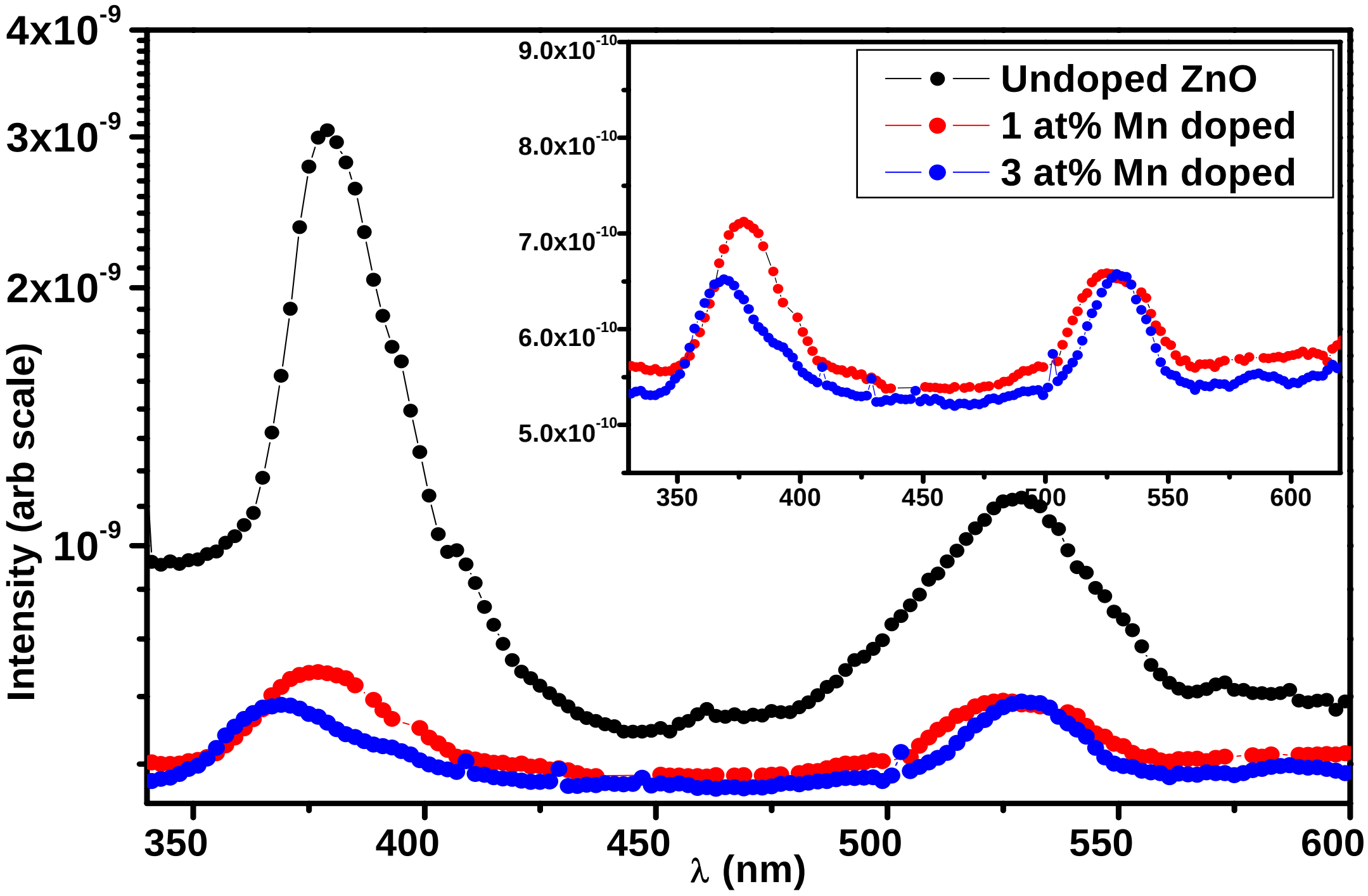

**Fig-6(a):** *Room temperature PL spectra of $Zn_{1-x}Mn_xO$ (x= 0.00, 0.01, 0.03)thin film samples. Inset shows intensity variation for 1 at% and 3 at% Mn doping.*

The interesting thing is that the intensity of NBE and broad emission in the higher wavelength region of the films becomes less intense with increasing Mn doping (shown in inset of Fig. 6(a)). And 5 at% Mn doped film (fig. 6(b)) didn't show any NBE emission around (~375 nm). The visible emission peak (peak around 525 nm) is assigned as due to the presence of singly occupied oxygen vacancy [31]. The ratio of intensities of NBE peak (peak around 375 nm) and VE peak indicate a progressive decrease as indicated in the inset of Fig. 6 (b). This is quite natural as with increasing Mn doping more and more defects have been incorporated in the system. So the NBE peak arising from real exciton recombination has been gradually suppressed by the increasing defect centre induced (singly occupied oxygen vacancy) recombination. Decrease in the ratio with the decrease in grain size predicts that visible emission center is located predominantly at the surface [32]. The reduction in emission intensity with Mn doping may be due to the increase in nonradiative recombination processes [33]. In presence of $Mn^{2+}$ ions in the film the emission intensity reduces considerably. Similar reduction effects of luminescence were reported in Mn and Co doped ZnO nanorods [34]. It might be due to the fact that the doped cations provide competitive pathways for recombination, which results in quenching of the emission intensity. So far as optical properties are concerned with increasing Mn doping it has been observed that more and more defect or disorder introduced in the system and accordingly the system has been greatly modified the optical properties.

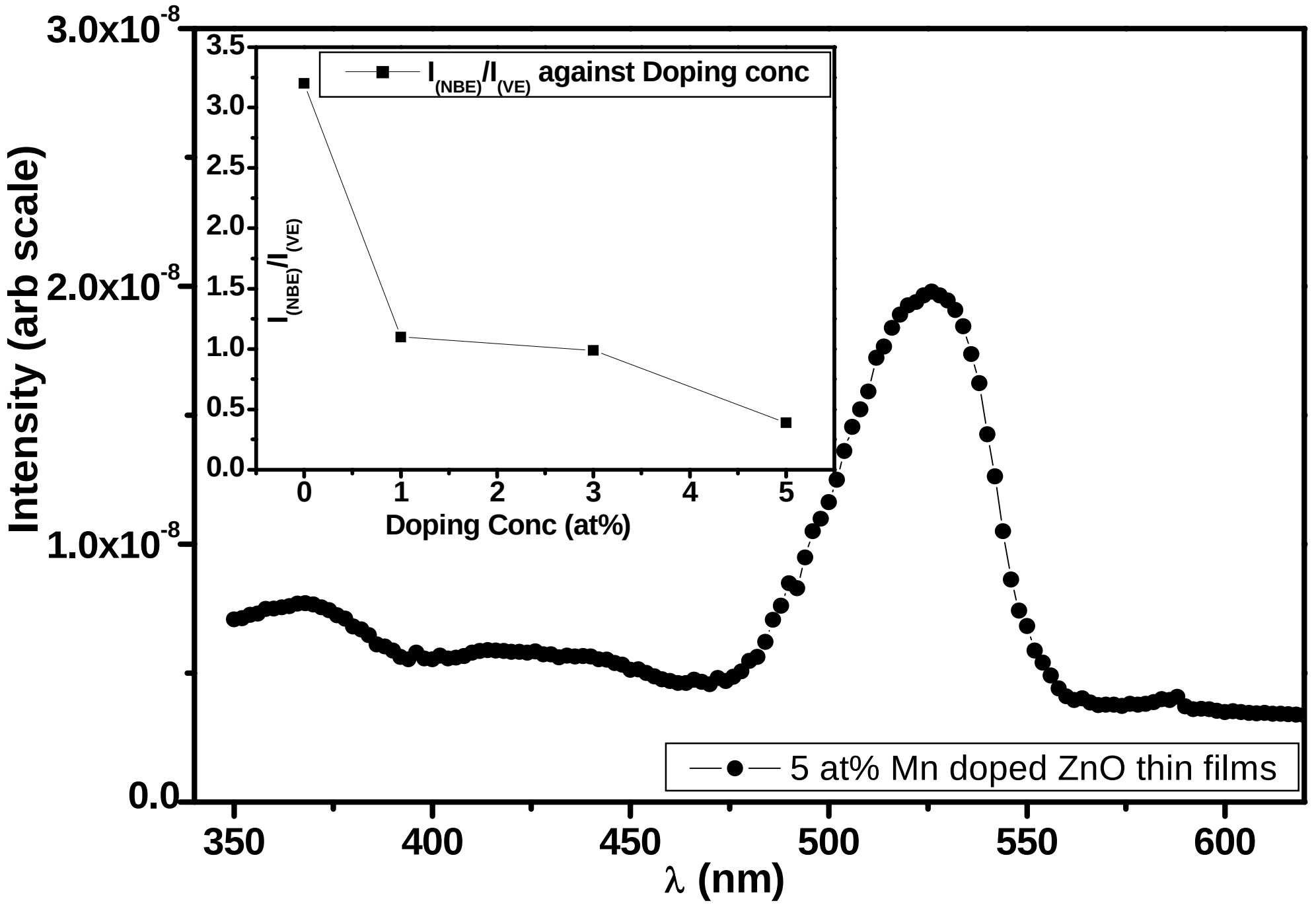

**Fig-6(b):** *PL spectra of $Zn_{0.95}Mn_{0.05}O$ thin film. Inset of (b) shows variation of ratio of PL intensities for NBE and defect related peaks with variation in Mn doping proportion.*

In order to investigate the magnetic property of $Zn_{0.95}Mn_{0.05}O$ thin film sample, magnetization measurement was performed by SQUID. The recorded M-H data had been subtracted from the diamagnetic contribution coming from the background signal of quartz substrate (shown in Fig (7)) using the expression [13, 35]

$$M_C = M_{F+S} - (-\Delta M/\Delta H)_S H \qquad (3)$$

Where $M_C$ is the magnetic moment after subtracting the substrate contribution, $M_{F+S}$ is the total magnetic moment of film and substrate and $(-\Delta M/\Delta H)_S$ H is the diamagnetic contribution due to the substrate. The Fig. 7 shows a hysteresis loop at 300K (room temperature) in the magnetic field with range ± 0.20T.

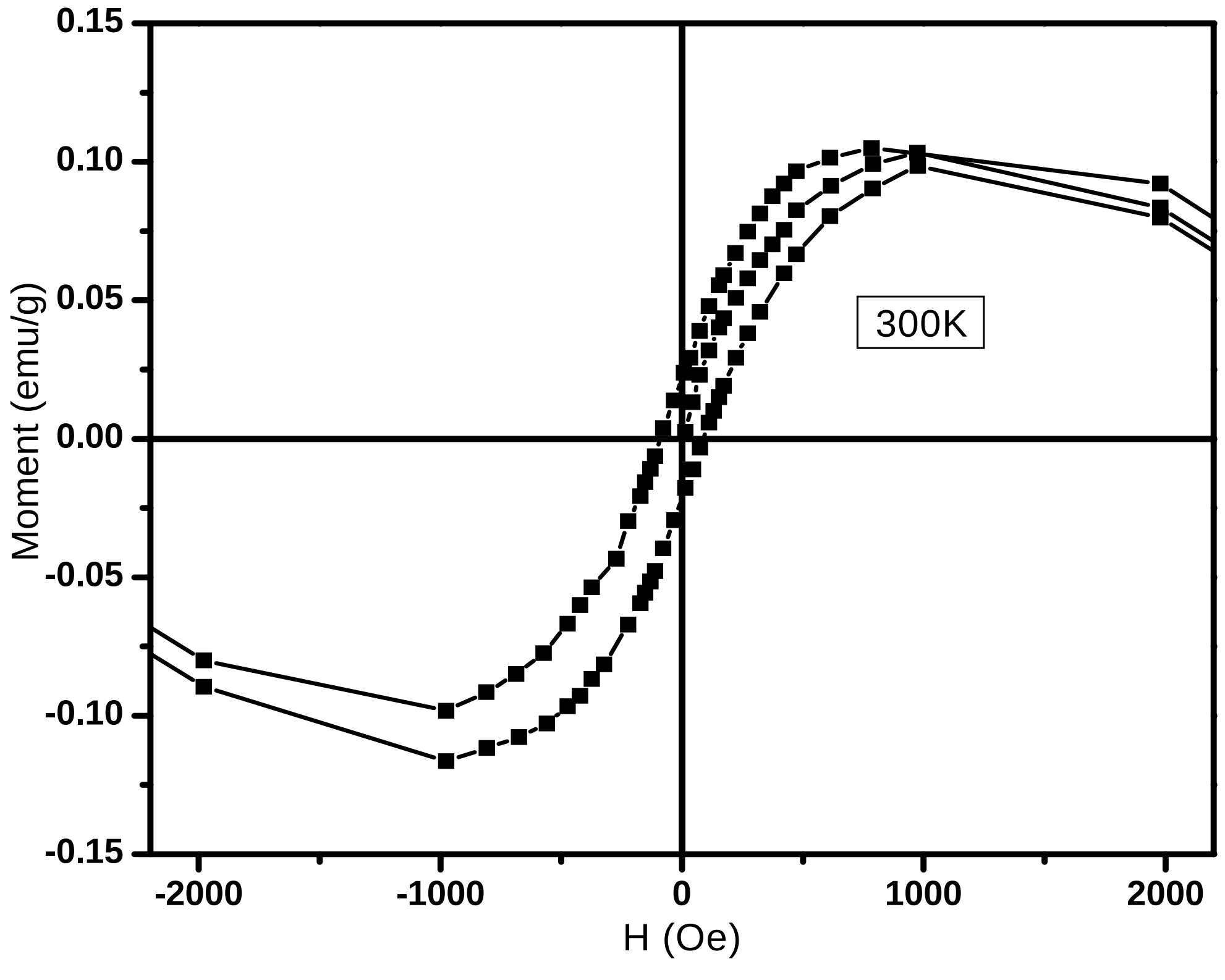

**Fig-7:** *Magnetization vs. magnetic field of $Zn_{0.95}Mn_{0.05}O$ thin film at 300K.*

Due to the presence of very minute amount of magnetic element in this kind of DMS systems ferromagnetism is always observed in the low field region. It is interesting to note that no trace of any Mn and/or Mn oxide cluster has been found in $Zn_{0.95}Mn_{0.05}O$ thin film sample as indicated from XRD and MFM measurements. So observed ferromagnetic nature of the sample at room temperature seems to be intrinsic. The calculated values of ferromagnetic saturation magnetization ($M_S$), remanence ($M_R$) and intrinsic coercivity ($H_c$) have been shown in Table 2. The value of saturation magnetization observed at room temperature as well as low temperature are much smaller compared to the theoretical value of $Mn^{+2}$ state (5 $\mu_B$/Mn atom) [36, 37]. For a free $Mn^{2+}$ ion with S=5/2 and g=2, suggests that the system poses antiferromagnetic (AFM) interaction along with ferromagnetic interactions between Mn ions [37]. Antiferromagnetic interaction arises from direct Mn-Mn interaction and thereby reducing the strength of ferromagnetic interaction. So magnetization of Mn ion is shown to be much smaller than predicted theoretical value.

Fig. 8 shows the temperature-dependent magnetization (M-T) of zero-field-cooled (ZFC) and field-cooled (FC) curve at an applied uniform external field of 200Oe and temperature range of 0-300K for $Zn_{0.95}Mn_{0.05}O$ thin film. Small bifurcation has been observed up to 100K for M(T) curves measured under FC and ZFC conditions.Both M(T) curves measured under FC and ZFC conditions indicate a peak around 56K. The possibility of presence of superparamagnetic nano particles in this system has been simply ruled out due to the following reason: in case of superparamagnetism M(T) curve under ZFC condition must show a gradual decrease in the low temperature regime with lowering in temperature (deblocking), while M(T) curve under FC condition continuously increase with reduction in temperature[38, 39]. So nature of bifurcation indicated by M(T) curves measured under FC and ZFC conditions from Fig. 8 confirms that the observed ferromagnetism is intrinsic in nature instead of any contamination of metallic nanoparticles assisted. The peak in both M(T) curves indicate a transition at temperature around 56K. This transition may be associated with the Mn induced defect incorporation in the system. It has been confirmed from XRD, AFM, optical measurements that Mn related defect states increases with increase of Mn doping. Therefore the observed ferromagnetism of the sample seems to be defect mediated.

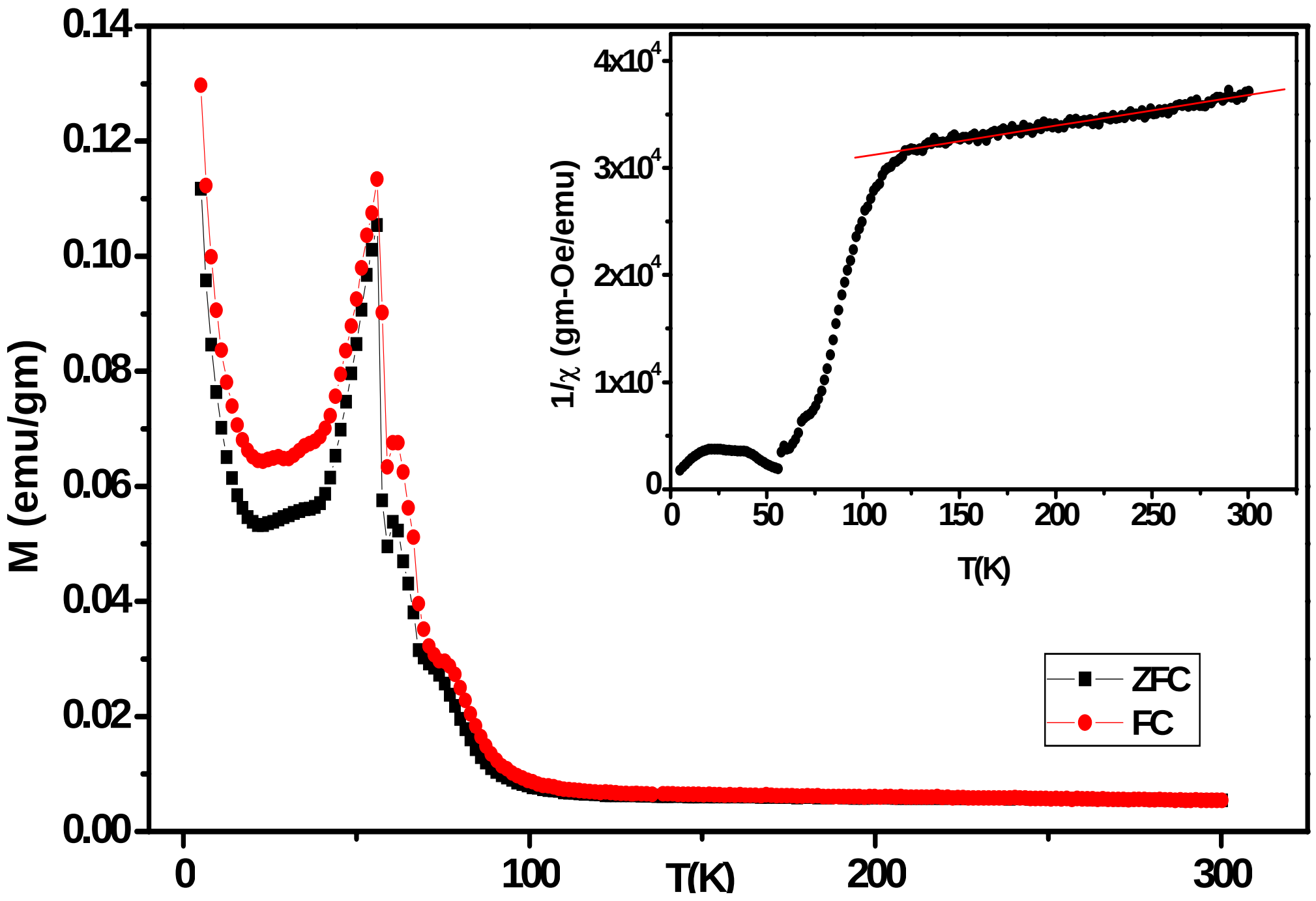

**Fig-8:** *Temperature dependence of magnetization of the $Zn_{0.95}Mn_{0.05}O$ thin film under a field of 200 Oe. Inset shows 1/x vs T curves of the $Zn_{0.95}Mn_{0.05}O$ thin film sample in a field of 200 Oe. The solid line is theoretical fits using the Curie–Weiss law.*

The temperature dependent inverse of dc magnetic susceptibility (χ) of $Zn_{0.95}Mn_{0.05}O$ thin film has been plotted in inset of Fig. 8. The 1/χ-T curve shows typical antiferromagnetic behavior. The straight line represents least square fit of Curie-Weiss law [$\chi = C(x)/\{T-\theta(x)\}$] to the experimental data in the temperature range 120 to 300K where x is the concentration of Mn atoms. The Curie constant $C(x)$ and Curie–Weiss temperature $\Theta(x)$ are expressed as

$$C(x) = xC_o = x(g\mu_B)^2 S(S+1)n/3k_B \quad (4)$$

$$\Theta(x) = x\Theta_0 = 2xS(S+1)z\, J/3k_B \quad (5)$$

where n is the no of Mn atoms/volume, g=2.0, $\mu_B$=9.27x10$^{-21}$ergs/Oe, and $k_B$=1.38x10$^{-16}$ ergs/K, and, S the value of the spin for the $Mn^{2+}$ ions in ZnO, $J$ is the effective exchange integral between nearest neighbor $Mn^{2+}$ ions and z is the no of nearest neighbor cations (z = 12 for wurzite structure). The value of $2J/k_B$ has also been estimated from equation (5) using S=5/2. The estimated values of C(x), $\Theta(x)$ and $2J/k_B$ is presented in Table 2.

**Table 2**: *Estimated magnetic parameters for $Zn_{0.95}Mn_{0.05}O$ thin film.*

| $C(x)$ | $\Theta(x)$ | $2J/k_B$ ($^oK$) | $H_{ci}$ (Oe) | $M^R_{FM}$ ($\mu_B$/Mn) | $M^S_{FM}$ ($\mu_B$/Mn) |
|---|---|---|---|---|---|
| 0.0347 | - 977.82 | 558.754 | 86.56 | 6.8636x10$^{-3}$ | 0.0298 |

The large and negative value of Θ(x) suggests presence of strong AFM interactions in the sample [13]. So the reason of observed low value of saturation magnetization of the sample has been sought qualitatively by estimating relatively high values of AFM interaction parameters.

## 3. Conclusions

$Zn_{1-x}Mn_xO$ (x=0.00, 0.01, 0.03, 0.05) thin films were synthesized by sol-gel technique. All of the films exhibit single phase nature. No trace of formation of any cluster of magnetic element Mn or its oxide(s) has been detected. The lattice parameters and unit cell volume increases with increasing Mn concentration. It indicates increase in Mn incorporation with increasing Mn doping. Grain size indicates decreasing trend with increasing Mn concentration. Band gap decreases and width of localized states ($E_0$) increases with increase in Mn doping. The photoluminescence study indicates increase of non-radiative transition with increase in Mn doping. Actually both these optical studies manifest increase in defect or disorder in the system with increasing Mn doping. 5 at% Mn doped ZnO film indicate intrinsic ferromagnetic nature at room temperature. Observed ferromagnetism seems to be defect mediated. The strong presence of AFM interaction has been established qualitatively and it actually acts as a reducing agent for the observed ferromagnetic moment of 5 at% Mn doped ZnO film.

## Acknowledgments

The work is financially supported by DST through project funding (vide project no.: SF/FTP/PS-31/2006). RK and SKN are thankful to the CRNN, C.U. and UGC, respectively for providing their research fellowship. We acknowledge UGC DAE CSR, Kolkata Centre, for utilizing their SQUID magnetometer facility. We also acknowledge DST for SPM facility prevalent in IUAC utilized for the AFM and MFM micrographs of the films under the IPHRA project. Further we are thankful to Dr. D. Kanjilal of IUAC to use their RBS measurement facility.